\documentclass[12pt,preprint]{aastex}

\shorttitle{Evolving Starburst Model for Neutral Gas Media}
\shortauthors{Yao}

\begin{document}

\title{Evolving Starburst Modeling of FIR/sub-mm/mm Line Emission. III. Application to Nearby Luminous Infrared Galaxies}

\author{Lihong Yao\altaffilmark{}}
\affil{Department of Astronomy and Astrophysics, University of Toronto,
    Toronto, ON M5S 3H8, Canada}
\email{yao@astro.utoronto.ca}


\begin{abstract}

In a previous work, we showed that the observed FIR/sub-mm/mm line spectra 
of a starburst galaxy (M 82) can be successfully modeled in terms of the 
evolutionary scheme of an ensemble of giant molecular clouds (GMCs) and shells, 
and such studies can usefully constrain the age(s) or star formation history 
of a starburst galaxy. In this paper we present a preliminary study of 
using the template of an ensemble of evolving GMCs/shells we developed 
for M 82. we apply the model to represent various stages of starburst evolution 
in a well known sample of nearby luminous infrared galaxies (LIRGs). 
In this way, we attempt to interpret the relationship between the degree 
of molecular excitation and ratio of far-infrared (FIR) to $^{12}$CO 
(or simply CO) luminosity to possibly reflect different stages of the 
evolution of star-forming activity within their nuclear regions.

\end{abstract}

\keywords{ISM: clouds -- ISM: evolution -- galaxies: ISM -- galaxies: starburst -- galaxies: star clusters -- radio lines: ISM}

\section{Introduction} \label{intro}

Star formation is one of the fundamental process that drives 
the evolution of galaxies. Understanding the interplay between star 
formation and the surrounding interstellar medium (ISM), as well as 
being able to parametrize the star formation history, are crucial in 
understanding the physics of galaxy evolution in the universe. 
To constrain theories of how the ISM evolves, one needs to investigate 
both individual galaxies and large statistical samples of data at 
multiple wavelengths. Especially, with the available data for the 
dust component, studying the gas in the co-space ISM becomes more 
interesting and important.

Starburst galaxies (e.g. M 82) have large reservoirs of molecular 
gas in their centers to fuel generations of star formation 
\citep[e.g.][]{wal02}. Bursts of massive star formation can create 
a profound impact on the surrounding ISM (both its structure and 
evolution) in their host galaxies by injecting large amounts of 
energy and mass into the ISM via strong stellar radiation, repeated 
supernova explosions, and cosmic rays \citep{rob98}. Hence, the main 
excitation mechanisms for molecular gas in starburst regions are 
the combination of collisional processes in dense gas and strong 
UV radiation field stemming from photoelectric heating, far-UV 
pumping of H$_2$, strong mechanical energy, enhanced cosmic rays 
and X-rays, as well as shock-induced and turbulent heating 
\citep[e.g.][and references therein]{mao00, vei09}. 

The volume of multi-wavelength observations of starburst galaxies 
throughout the cosmic-scale has increased dramatically due to the 
significant improvement in the sensitivity and resolution of available
instrumentation. These observations provide an essential basis for 
starburst modeling; in turn, models provide systematic predictions 
of the properties of the ISM in starburst galaxies. In two previous 
papers (Yao et al. 2006, hereafter Paper I; Yao 2009a, hereafter 
Paper II), we presented a unique set of starburst models that allow 
us for the {\it first time} to relate observed molecular line 
properties of a starburst region to its age and provided better 
interpretations for the observations. Few previous models for 
neutral gas media, if any, have all physical elements included 
at the same time to search for the time signal in the CO spectral 
energy distribution. In our model, the molecular gas responding 
to star formation in a starburst galaxy is treated as an ensemble 
of GMCs with different initial masses, each of which responds 
to massive star formation at its center. The four physical elements 
incorporated in our model are i) a standard similarity solution 
for the bubble/shell structure around a super star cluster (SSC), 
ii) a time-dependent stellar population synthesis method, iii) a fully 
time-dependent chemistry problem for the PDRs, and iv) a non-LTE 
radiative transfer method for molecular and atomic lines. Detailed 
discussion of each of these elements and the importance in evaluating 
the physical and chemical properties of molecular gas in a starburst 
galaxy, are presented in the Ph.D. thesis of Yao (2009b, and 
references therein, hereafter Yao Thesis\footnote{The thesis is available 
from the University of Toronto library website: 
http://www.astro.utoronto.ca/AALibrary/theses.html}). 

In Papers I and II, we have successfully demonstrated that the 
kinematic and far-infrared/sub-millimeter/millimeter (FIR/sub-mm/mm)
emission properties of individual expanding shells and star-forming 
regions in nearby archetype starburst galaxy M 82 can be understood 
by following the evolution of individual SSC or an ensemble of SSCs 
surrounded by compressed shells and GMCs \citep{ket05, mel05, wei05, mcc07}. 
We found that the spectral energy distributions (SEDs) of the molecular 
and atomic line emission from these swept-up shells and the associated 
parent GMCs contain a {\it signature} of the stage of the starburst 
evolution. We suggested that the physical and chemical properties 
of molecular and atomic gas and their excitation conditions 
depend strongly on the stage of starburst evolution in the central
kiloparsec region of M 82. 

At distances beyond M 82 (3.25 Mpc; Lord et al. 1996), the sub-millimeter 
common user bolometer array (SCUBA) sources represent an inherently 
distinct population from low-$z$ to high-$z$ universe \citep{hol99, san99}. 
Multi-wavelength studies of these galaxies in the local universe, 
for example, the SCUBA local universe galaxy survey 
\citep[SLUGS;][]{dun00, tho02, yao03}, has revealed that the properties 
of molecular gas and dust in the central starburst regions differ 
significantly from those of quiescent star forming disks and the 
center of the Galaxy. The SLUGS dust sub-millimeter survey \citep{dun00} 
contains 104 galaxies selected from the {\it Infrared Astronomical 
Satellite} (IRAS) bright galaxy sample, with a flux limit of 
$S_{60}$ $>$ 5.24 Jy, and FIR luminosities in the range 
10$^{10}$ L$_{\odot}$ $<$ $L_{FIR}$ $<$ 10$^{12}$ L$_{\odot}$, 
and the SLUGS CO survey \citep{yao03} contains about 60 SLUGS galaxies 
over a restricted range in RA. It is essentially complete to the same 
{\it IRAS} flux based on the selection criteria of a flux-density and 
distance limited for the SLUGS sample.

Low-lying CO rotational line transition at millimeter wavelength 
(CO $J$ = 1 - 0) is often used as a tracer of the total molecular 
hydrogen content in a galaxy \citep[e.g.][]{mau96, mau99, yao03}.
But it is not a good tracer of dense or warm gas in star-forming 
regions. The most common method of deriving H$_2$ masses from 
$^{12}$CO(1-0) luminosities is through the controversial parameter 
$X$, which converts CO line intensity or luminosity to the H$_2$ 
column density or mass. Previous studies have shown that this 
parameter varies from galaxy to galaxy \citep{bna98, bos02, zhu09}, 
and it is thought to be higher in metal-poor galaxies and lower 
in starburst galaxies than in Galactic molecular clouds, where 
$X$ is about 2.8 $\times$ 10$^{20}$ cm$^{-2}$ [K km s$^{-1}$]$^{-1}$ 
\citep{blo86, str88}. Hereafter we refer to this as the standard 
value. Thus in starburst galaxies, application of the standard factor 
can produce a significant overestimate (4 - 10 times) of molecular 
hydrogen mass \citep{sol97, dns98, yao03}. The question we ask 
is {\it Does the $X$-factor depend on the phase of starburst 
evolution?}.

In \S~\ref{slugs} of this paper we present the application 
of our evolving starburst models to the SLUGS sample, using the 
template of a shell ensemble we developed for M 82 (Paper II). 
We examine one interesting and yet puzzling issue in particular 
- whether the relation between the degree of molecular gas excitation 
and the star formation properties in these SLUGS objects can be 
understood in terms of our evolving starburst model as a preliminary 
study. This issue is a follow-up to an earlier observational paper 
by Yao et al. (2003) in which a clear connection between these 
two properties was identified. We also investigate whether the 
variations in these properties from galaxy to galaxy may 
derive simply from seeing galaxies in different stages of their post 
starburst evolution, as seen in our models. Previous interpretation 
of these effects require that they reflect the diversity in the 
intrinsic properties of galaxies, with no necessary connection to 
starburst evolutionary phases. In \S~\ref{xfact} of this paper, 
we present a {\it first attempt} to model the time dependence of 
this well known controversial CO-to-H$_2$ conversion factor $X$ 
in starburst regions.  

\section{Correlation between CO Excitation and Star Formation Properties} \label{slugs}
 
We first compute the relevant star formation related characteristics 
(e.g. degree of molecular gas excitation, star formation efficiency) 
as a function of time (i.e. the stage of starburst evolution) for 
a model shell ensemble, then we compare these predicted characteristics 
with those observed in a modest sample of nearby luminous IR galaxies. 
Each galaxy in the sample is presumed to represent a different 
evolutionary stage of our model starburst. We specifically investigate
the ratios of quantities, $^{12}$CO(3-2) to (1-0) line intensity 
ratio $r_{31}$ \citep{yao03}, and the FIR luminosity 
(10 $<$ $\lambda$ $<$ 100 $\mu$m) to molecular gas mass ratio 
$L_{FIR}$ / $M$(H$_2$), that are used in our analysis. These ratios 
describe intrinsic properties independent of galaxy size and assumed 
distance \citep{yao03}. The lack of such a correlation may be due 
to a sample selection effect, since the more distant galaxies
may have a larger fraction of highly excited CO molecular gas.
In the standard literature, the $L_{FIR}$ is used as an indicator 
of star formation rate \citep[SFR;][]{ken98}, and the quantity 
$L_{FIR}$ / $L_{CO}$ or $L_{FIR}$ / $M$(H$_2$) is taken to be 
a measure of the star formation efficiency \citep[SFE;][]{yao03}, 
i.e. star formation rate per unit available molecular gas mass. 
However, in our model this ratio does not simply measure a constant 
star formation efficiency, but is a parameter which undergoes a dramatic 
evolution during the tens of millions of years following a starburst event. 
The FIR luminosity $L_{FIR}$ can be readily derived from a dusty starburst 
model \citep{efs00}, but currently our starburst model is not available
for computing the continuum emission from the dust media. In this paper, 
we use the stellar cluster luminosity $L_{SC}$ to represent the $L_{FIR}$, 
by crudely assuming that all of the UV photons is absorbed by dust enshrouding 
the star clusters and distributed in the galaxies, processed and re-emitted 
in the far-infrared. The ratio of $^{12}$CO(3-2) to (1-0) line emission 
$r_{31}$ is known to provide a more sensitive measure of the gas 
temperature and density than the ratio of $^{12}$CO(2-1) to (1-0) 
lines \citep[e.g.][]{mau99}. The $r_{31}$ ratio is often used to 
measure the degree of molecular CO excitation \citep{san93, yao03}.  

An important feature observed in luminous IR galaxies is that 
the degree of $^{12}$CO excitation tends to increase with 
increasing concentration and efficiency of star formation activity
\citep{yao03, wil08}. Here we examine the effects of starburst phase in our 
model on the excitation ratio $r_{31}$ by comparing a theoretical plot 
based on our modeling results with this observed relationship found 
in Yao et al. (2003). We begin with the explanation of the effect outlined 
by Yao et al. (2003), and follow this with a different possible origin based 
on our starburst evolution scenario.

\begin{figure}
\epsscale{1.0}
\plotone{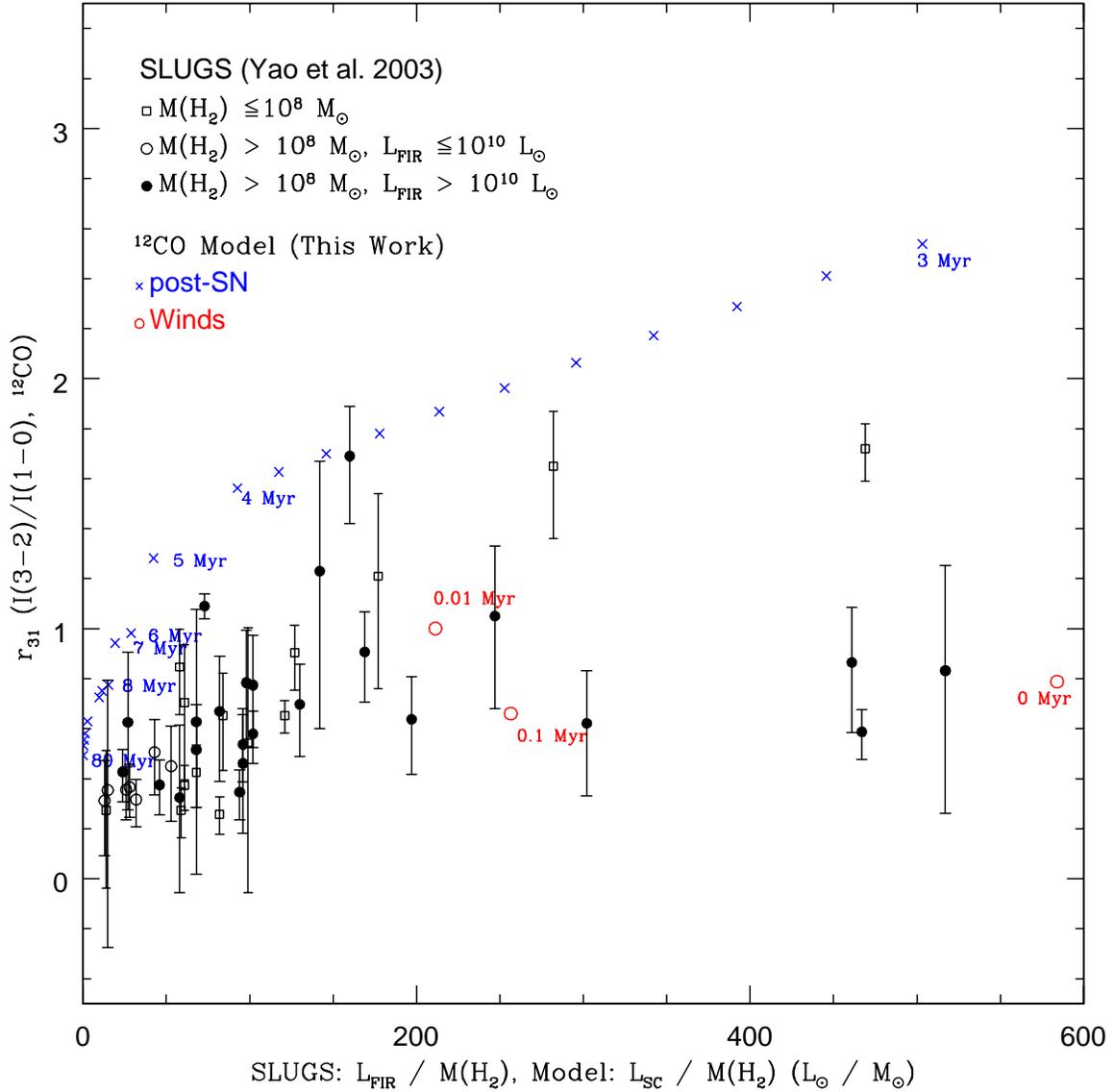}
\caption[Line intensity ratio $r_{31}$ versus $L_{FIR}$/$M$(H$_2$) ratio]
{Plot of the line intensity ratio $r_{31}$ (expressed as brightness temperatures 
integrated over velocity) versus the $L_{FIR}$/$M$(H$_2$) ratio. The symbols 
with error bars are the observed data $r_{31}$ measured within a 15$^{\prime\prime}$ 
aperture for the SLUGS sample. The uncertainties for the $L_{FIR}$ data are not 
available, therefore there is no horizontal error bar shown in the plot. Also shown 
in the plot are our model results, red open symbols (square: $t$ $<$ 0.01 Myr, 
circle: 0.01 $\le$ $t$ $\le$ 0.1 Myr, triangle: 0.1 - 0.7 Myr) are for the {\it Winds} phase, 
and blue crosses are for the {\it post-SN} phase. The model age sequence for the 
{\it post-SN} phase from right to left is 3, 4, 5,$\ldots$, 80 Myr.
\label{r2sfe}}
\end{figure}

Fig.~\ref{r2sfe} (see also Fig. 10 of Yao et al. (2003)) 
shows that there is a significant observed correlation between 
$r_{31}$ and $L_{FIR}$/$M$(H$_2$) within the 15$^{\arcsec}$ aperture
for the data with $L_{FIR}$/$M$(H$_2$) $\le$ 200 L$_{\odot}$/M$_{\odot}$.
The SLUGS sample shown is divided into two ranges by gas mass 
centered at $M$(H$_2$) = 10$^8$ M$_{\odot}$, and also by dust 
FIR luminosity centered at $L_{FIR}$ = 10$^{10}$ L$_{\odot}$ 
which are indicated by three different symbols in the plot. 
The segmentation according to gas mass range and dust IR 
luminosity range shows the relationship between $L_{FIR}$ and 
$M$(H$_2$) and position in the plot. The linear correlation 
coefficient is 0.74 at a significance level of 1.4 $\times$ 10$^{-7}$ 
(i.e. probability that $r_{31}$ and $SFE$ are related is significant).
However, the correlation is diminished at $L_{FIR}$/$M$(H$_2$) $>$ 
200 L$_{\odot}$/M$_{\odot}$, where $r_{31}$ ranges  between 
0.5 and 1.72 \citep{yao03}. The molecular gas mass in the SLUGS 
sample is derived from CO luminosities by applying the conversion 
factor $X$ = 2.7 $\times$ 10$^{19}$ cm$^{-2}$ [K km s$^{-1}$]$^{-1}$ 
estimated using an large velocity gradient (LVG) analysis \citep{yao03}, 
which is about 4-10 times lower than the conventional $X$ 
value derived from the Galaxy \citep{kny89}. Hence, the result is 
$M$(H$_2$) = 1.1 $\times$ 10$^3$ $D^2_{L}$/(1+$z$) $S_{CO}$ M$_{\odot}$, 
where $D_{L}$ is the luminosity distance of a galaxy in Mpc, 
and $S_{CO}$ is the $^{12}$CO(1-0) flux in Jy km s$^{-1}$ measured 
within a 15$^{\prime\prime}$ beam. According to Yao et al. (2003), 
both $r_{31}$ and $L_{FIR}$/$M$(H$_2$) ratios are found to be 
independent of the galaxy distance (or the projected beam size on 
galaxy), and in turn the H$_2$ gas mass. There are also no 
significant correlations found between $r_{31}$ and star 
formation rate (or $L_{FIR}$), dust temperature and mass, the 
color indices, or the luminosity of the IR or radio continuum. 
Yao et al.  (2003) suggested that the observed correlation between $r_{31}$
and $L_{FIR}$/$M$(H$_2$) and the lack of correlation of $r_{31}$ 
with properties related to total star formation imply a dependence 
only on localized conditions within the molecular clouds. According to this 
picture, the dependence of $r_{31}$ on the $L_{FIR}$/$M$(H$_2$)
ratio reflects a higher degree of CO excitation that is associated 
with a higher spatial concentration and efficiency of star forming 
activity. Such conditions would arise in an intense starburst where 
the surface density of such activity is high. The saturation effect
(approaching unity) of $r_{31}$ seen at $L_{FIR}$/$M$(H$_2$) $>$ 200 
L$_{\odot}$/M$_{\odot}$ reflects a limit imposed on this ratio 
at the highest excitation where the excitation temperatures for 
$^{12}$CO(3-2) and (1-0) (both assumed optically thick) are equivalent.  

Starbursts remain a viable explanation for the dominant energy 
source in luminous infrared galaxies at the local universe. 
An instantaneous starburst model is therefore assumed here for 
the studies of the SLUGS galaxies, the same as our previous studies 
of nearby starburst galaxy M 82 (Paper II). The instantaneous burst 
scenario may not be physically realistic for SLUGS galaxies, but it 
is an acceptable representation of the molecular CO line SED if the 
duration of the star-forming event in their centers is shorter than 
the starburst age (Paper II). The set of our starburst models is 
divided into two phases defined in Paper II, the {\it Winds} 
phase comprises younger and denser shells than in the {\it post-SN}
phase. These two phases are treated independently, and can be viewed 
as simply alternative scenarios.  

As mentioned earlier the $L_{FIR}$/$M$(H$_2$) ratio is traditionally 
used as an indicator of star formation efficiency. In our model,  this ratio 
and its variation with time are simple and direct consequences of the 
evolution in stellar luminosity (represented by $L_{SC}$) and swept-up 
gas mass within a single starburst. Fig.~\ref{r2sfe} also shows the model 
$r_{31}$ ratio versus the $L_{SC}$/$M$(H$_2$) ratio, where the total cluster 
luminosity $L_{SC}$ is the sum of individual SC stellar luminosities in
our model shell ensemble (see Table 1 in Paper II). The $L_{SC}$  has been 
used in place of the observed FIR luminosity $L_{FIR}$. This assumes that the 
FIR luminosity produced in a recent starburst is the dominant component 
of the FIR luminosity.  The $M$(H$_2$) for the model is derived from $S_{CO}$, 
thus using the same method as employed for the SLUGS sample except with 
a fixed $D_L$ = 3.25 Mpc used in our M 82 model computations (Paper II). 
Thus, $S_{CO}$ and $M$(H$_2$) are both functions of time and the size of the 
shells. The model curve clearly shows a trend similar to that observed 
$r_{31}$ ratio increases with increasing $L_{SC}$/$M$(H$_2$) 
ratio, and then saturates at high 
$L_{SC}$/$M$(H$_2$) ($>$ 200 L$_{\odot}$/M$_{\odot}$). The model 
$r_{31}$ ratios vary between 0.7 and 1.3 for the {\it Winds} phase, 
and between 0.4 and $>$ 2.0 for the {\it post-SN} phase. The key 
point associated with our model results shown in Fig.~\ref{r2sfe} 
is that the relationship between $r_{31}$ and $L_{SC}$ is governed 
by the age or phase of the starburst. At earlier stages the cluster 
luminosity is high and the mass of swept-up gas is small. Since the 
shells are comparatively small and expanding rapidly, they are also 
more effectively heated and compressed than at later stages. Thus, at 
earlier phases, both $r_{31}$ and $L_{SC}$/$M$(H$_2$) are higher 
than in the later phases of the expansion. Since the parent GMCs 
contribute more to the $^{12}$CO(1-0) line than the (3-2) line, the 
$r_{31}$ ratio is lower for the {\it Winds} phase, although the $L_{SC}$ 
is higher than that from the {\it post-SN} phase. An important point 
associated with this interpretation is that the degree of molecular gas 
excitation is a consequence of star-forming activity, rather than a reflection 
of initial conditions prior to the starburst, as often assumed. The physical and 
chemical properties, as well as the structure of the ISM, which are often 
traced by different molecular species in a starburst regions, have been 
modified significantly throughout the evolution of the starburst.  Hence,
higher degree of molecular gas excitation does not necessary reflect, 
for example, higher star formation efficiency.

As described above, our model yields a result similar to the observations 
(i.e. possibly an upper limit of the observed data) and suggests that the 
observed behavior results from recent starbursts in these infrared 
luminous galaxies. The ratios predicted by our model are higher than
the observed SLUGS data, because (1) we use a smaller region (possibly 
with more highly excited gas) in the model,  (2) the model CO(1-0) line flux 
may be underestimated due to the limitations of the model (e.g. neglecting 
the emission from the low-density ambient ISM) discussed in Paper II, 
and (3) we use the total stellar luminosity for the total FIR luminosity in
the model, although the total stellar luminosity derived from our model 
may be underestimated also discussed in Paper II.  This result 
furthermore implies that the relationship between the degree of CO 
excitation ($r_{31}$) and the $L_{FIR}$/$M$(H$_2$) ratio associated 
with star formation properties may be determined by the phase of the 
starburst rather than by the more traditional view of variation in the 
efficiency of star formation. 

\begin{figure}
\epsscale{1.0}
\plottwo{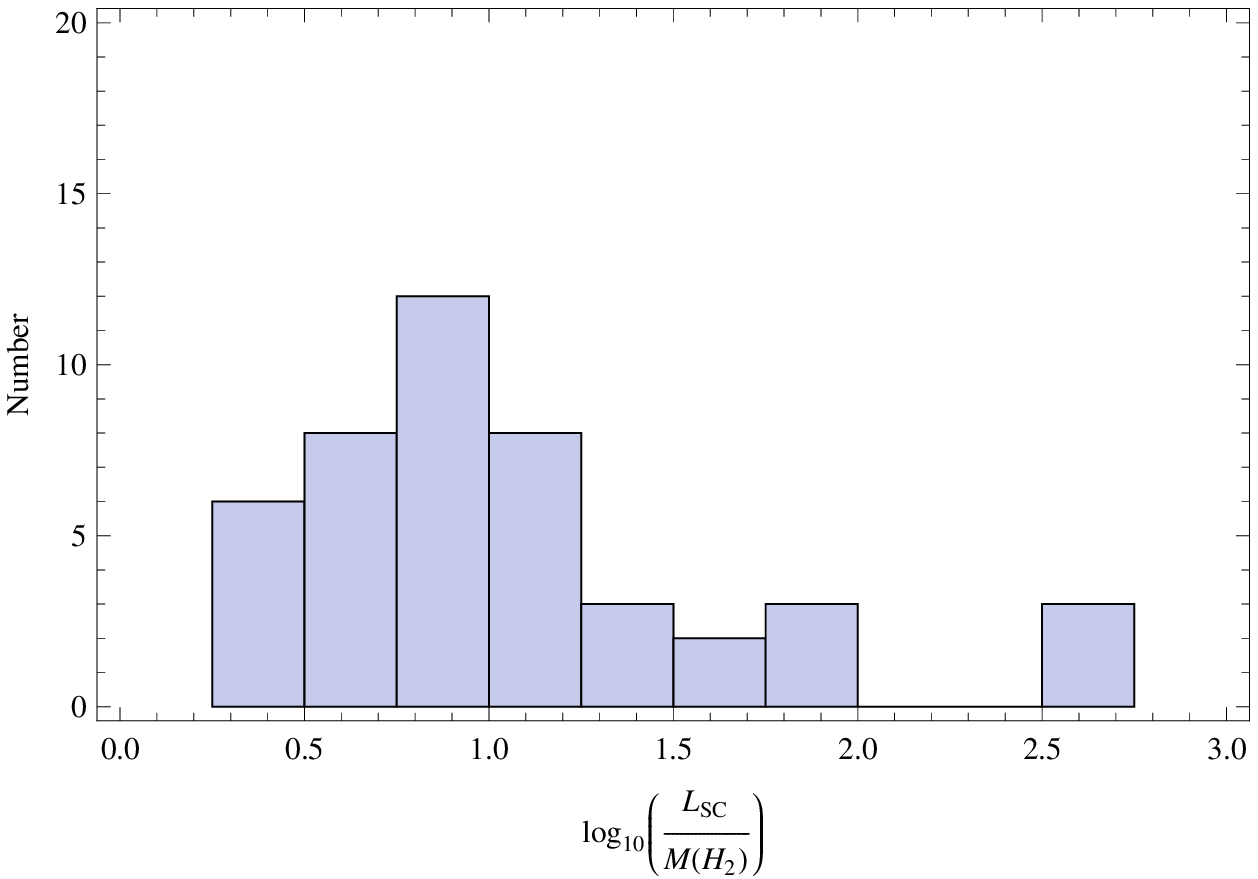}{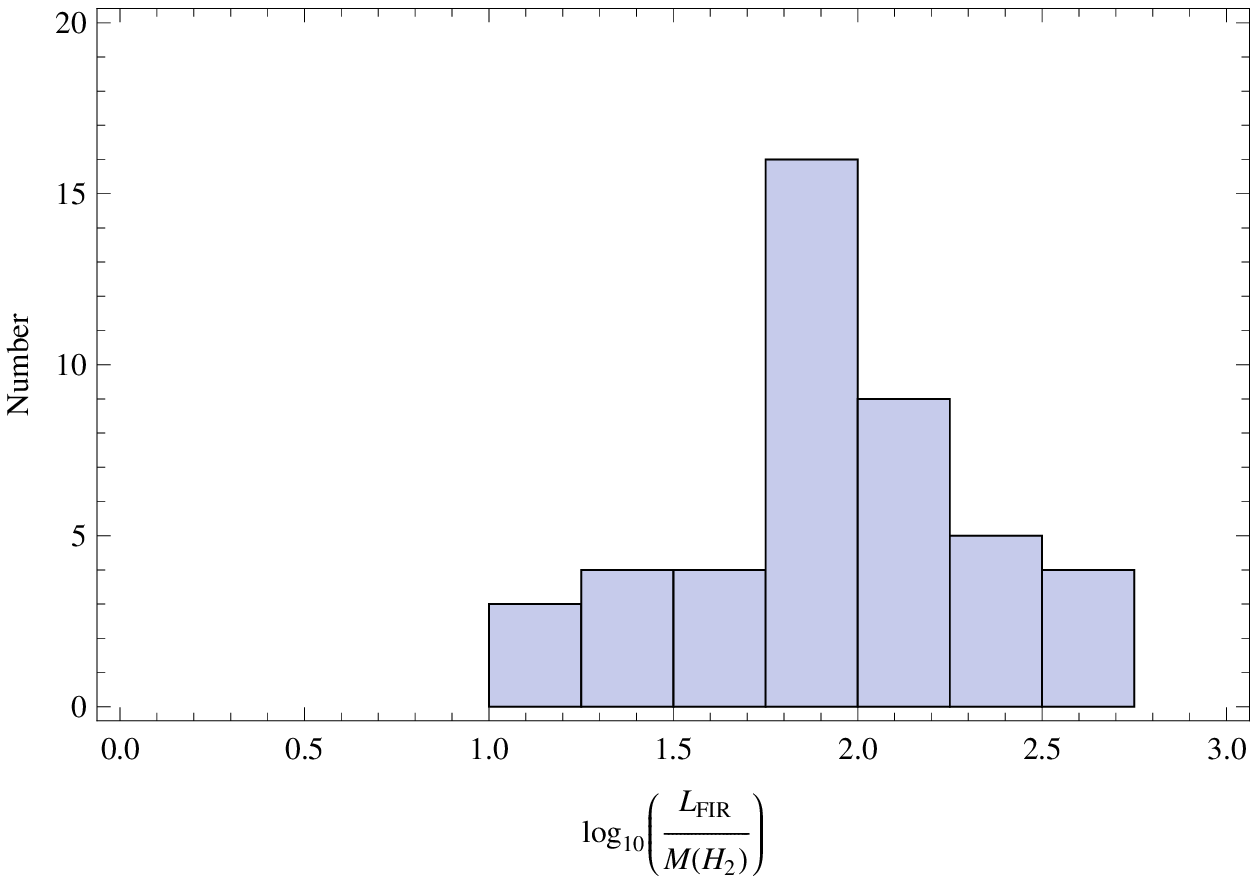}
\caption[Histograms of our model $L_{SC}$/$M$(H$_2$) ratio and observed 
$L_{FIR}$/$M$(H$_2$) ratio in SLUGS galaxies]
{Histograms of the $L_{SC}$/$M$(H$_2$) ratio derived from our starburst 
model and the $L_{FIR}$/$M$(H$_2$) ratio measured for 45 SLUGS galaxies 
by Yao et al. (2003). The  model ages are between 3 Myr and 20 Myr. 
The comparison between two histograms show qualitative similarity. 
The difference in the location of the peak may be caused by the deficiency 
in the model luminosity (Paper II). 
\label{sfehist}}
\end{figure}

\begin{figure}
\epsscale{1.0}
\plottwo{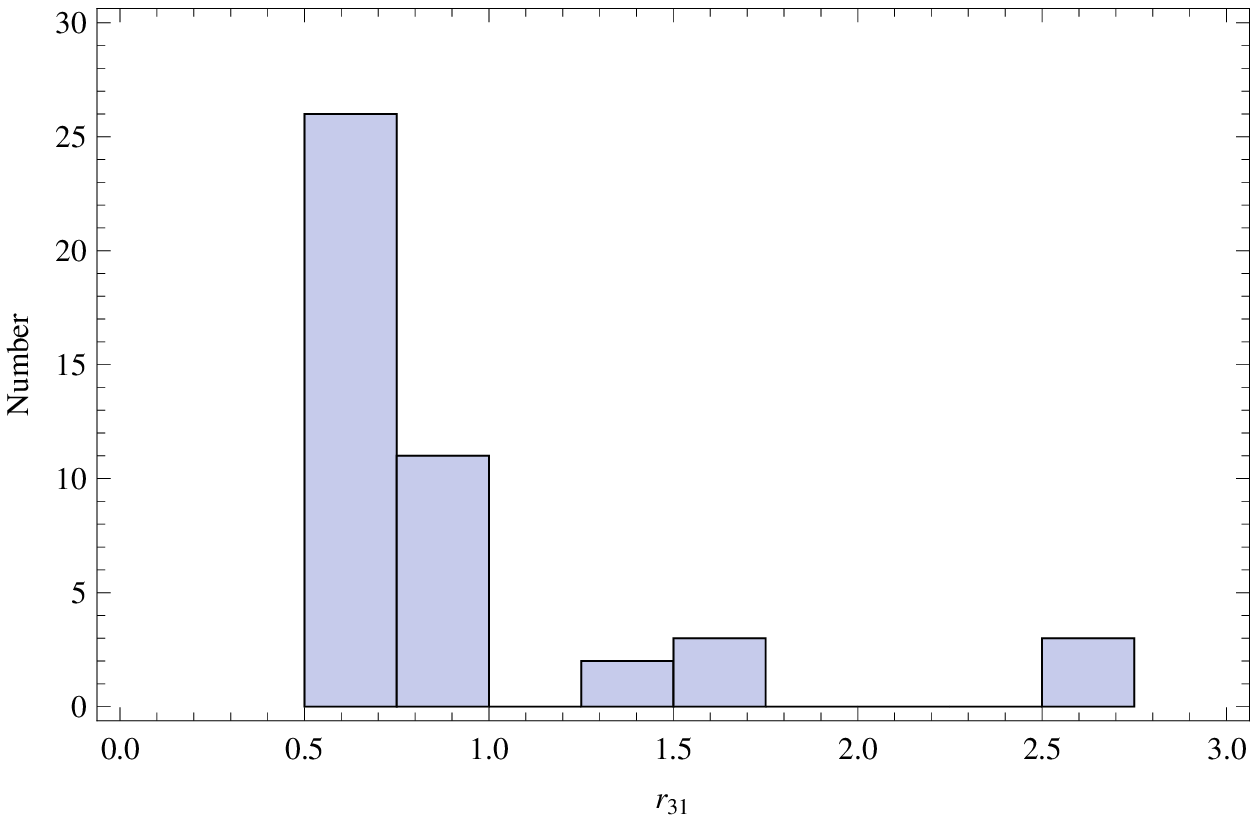}{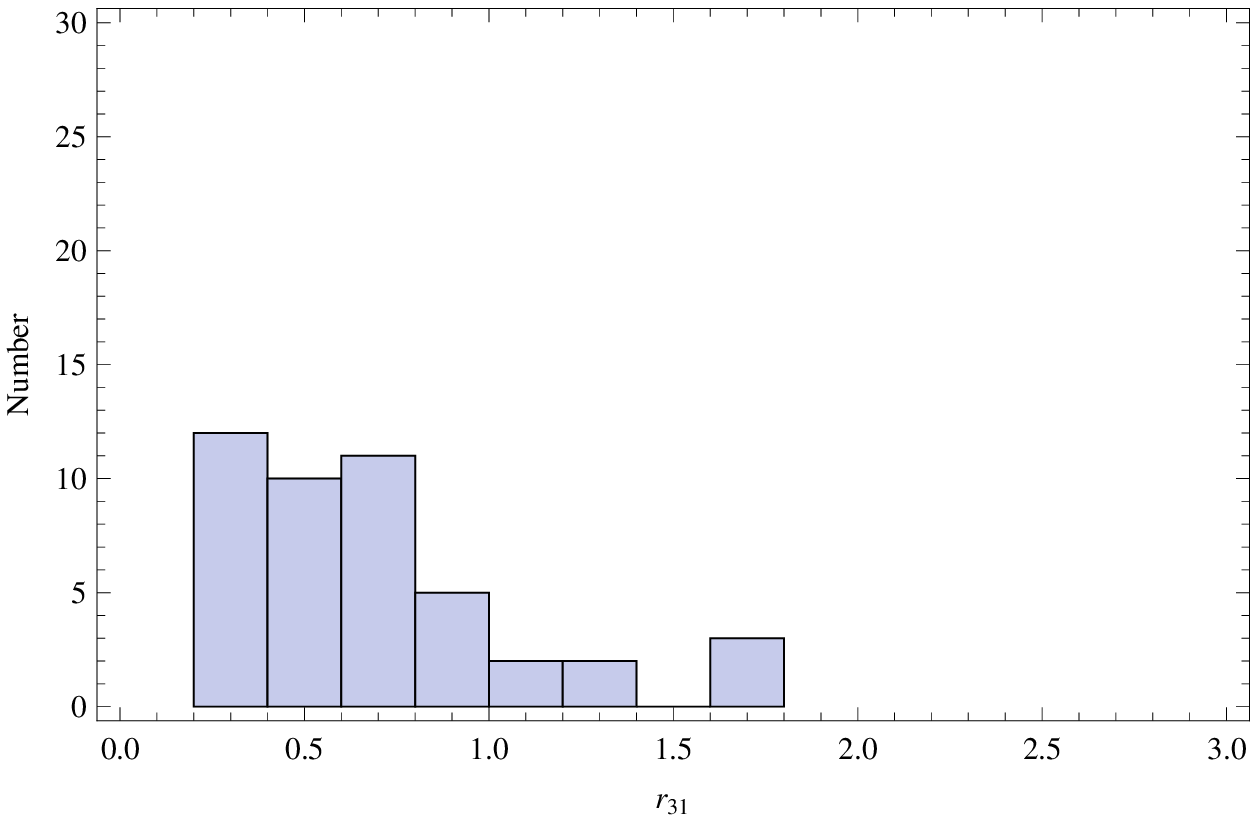}
\caption[Histograms of our model $r_{31}$ ratio and observed $r_{31}$ 
ratio in SLUGS galaxies]
{Histograms of the $^{12}$CO(3-2)/(1-0) line ratio ($r_{31}$) derived from 
our starburst model and the $r_{31}$ ratio measured for 45 SLUGS galaxies 
by Yao et al. (2003). The  model ages are between 3 Myr and 20 Myr. 
The comparison between two histograms show qualitative similarity. 

\label{r31hist}}
\end{figure}

One of the interesting properties of Fig.~\ref{r2sfe} is that the observed 
data points tend to concentrate toward the origin. A similar effect 
is discernible in the theoretical plot if points are plotted at equal 
intervals in age. The concentration in the latter case occurs 
because the rate of change of the variables on both axes decreases 
with time as the age becomes large. This suggests a further test of 
the hypothesis that the observed behavior in Fig.~\ref{r2sfe} is the 
consequence of seeing starbursts in different stages of their evolution. 
Accordingly, we compare the distribution of the 45 SLUGS galaxies with 
respect to the observed $L_{FIR}$/$M$(H$_2$) and $r_{31}$ ratios to that 
of 45 pseudo galaxies using the model $L_{SC}$/$M$(H$_2$) and $r_{31}$ 
ratios, as shown in Figs.~\ref{sfehist} and~\ref{r31hist}. The starbursts in the 
pseudo galaxies were assigned ages drawn randomly from a uniform 
probability distribution between 3 Myr and 20 Myr, and the parameters 
$r_{31}$ and $L_{SC}$/$M$(H$_2$) then computed for these ages using our 
model. The model age range (3 - 20 Myr) reflects the range of validity of 
the model for the {\it post-SN} phase. The lower boundary corresponds 
to a plausible lower limit on the dynamical timescale for the starburst 
region and the upper boundary corresponds to the epoch beyond which 
the bubble shells escape the disk of the galaxy. The comparison 
shown in Figs. 3 and 4 reveals a similarity in the distributions between 
model and observation for both log$_{10}$ ($L_{FIR}$/$M$(H$_2$)) and 
$r_{31}$, though the peak in the former distribution occurs at a lower 
value in the model (a point to be discussed further later in this session). 
The qualitative similarity between the two histograms thus supports 
the hypothesis that the quantities $L_{FIR}$/$M$(H$_2$) and $r_{31}$ 
are related to starburst age. However, the evidence presented is not 
conclusive, merely suggestive.

We also tested our model result with different starburst age ranges 
(1 to 10 Myr and 1 to 80 Myr) for the 45 pseudo galaxies. These tests 
produced numerically different but qualitatively similar results.

As noted, the analysis encapsulated in Fig.~\ref{r2sfe} signifies that the 
evolution of gas excitation properties during the cause of a starburst.
However, there are a number of considerations which need to be 
examined which may affect the credibility of this result, e.g., the selection 
effect on the observed frequency distributions. Fortunately, the SLUGS 
subsample investigated here is nearly complete with a limiting FIR flux 
density and a limiting distance, so that the selection effects are well understood. 
Virtually all members of the sample are detected at both CO transitions, 
but the flux limit imposes a minimum detectable luminosity which 
increases with $D^2_L$. The dramatic decline in galaxy number 
density with distance confirms this selection and indicates that 
the sample comprises the high luminosity tail of the underlying 
galaxy population. The question then is: could the luminosity selection 
affect the distribution of $L_{FIR}$/$M$(H$_2$), particularly in producing 
a deficiency of ratios below the peak of the observed distribution? 
There are two approaches to investigate this effect. First, we can examine 
the direct relations between $L_{FIR}$/$M$(H$_2$) and galaxy distance, 
as well as $r_{31}$ and galaxy distance (see Figs. 3 and 5 in Yao et al. 2003). 
We find no significant correlation between these two quantities and  
distance. Second, we divide the sample of 45 SLUGS galaxies used in 
our frequency distribution tests (Figs. 3 and 4) into two parts, each with 
23 objects, divided at $D_L$ = 45 Mpc. We find no significant difference 
between the frequency distributions of $L_{FIR}$/$M$(H$_2$) in these 
two subgroups. Thus, there is no evidence that the selection effect in 
luminosity produces a corresponding selection in the ratio 
$L_{FIR}$/$M$(H$_2$). We also conducted a similar analysis for the ratio 
$r_{31}$,  and there is also no evidence of selection effect on this ratio.   

Other observational effects, such as, the random and systematic errors 
in the observed data, would contribute to the disagreement between 
the model and the observed histograms (i.e. Figs. 3 and 4), assuming 
that the theory were the correct explanation for the observation. It must 
also be recalled here that there is a deficiency of about an order of magnitude 
between model and observation in the total stellar luminosity for M 82. 
This deficiency in the model luminosity will contribute to, and possibly 
even account for, the systematic difference in location of the peak of the 
two distributions of $L_{FIR}$/$M$(H$_2$) shown in Fig.~\ref{sfehist}.     

The hypothesis presented here that the ratios $r_{31}$ and 
$L_{FIR}$/$M$(H$_2$) are related to the starburst age can be 
further tested by direct measurements of the ages of the young 
stellar populations in the SLUGS subsample. Probably the best 
approach would be the fitting of population synthesis models to 
optical and IR spectroscopy of SLUGS objects, as discussed in 
Paper II.

\clearpage

\clearpage

\section{The CO-to-H$_2$ Conversion Factor $X$} \label{xfact}

Our evolving starburst model allows us to investigate, purely from 
a theoretical standpoint, the relationship between the $X$-factor 
and starburst phase, because the physical properties of molecular 
gas in an evolving starburst region changes with time. The 
$X$-factor may be determined from the following equation \citep{yao03},

\begin{equation}
X(t) = \frac{M(H_2)}{4.1 \times 10^2 D^2_{L} S_{CO}}
\end{equation}

where $M$(H$_2$) is the total H$_2$ gas mass swept up by the shells 
at time $t$ in units of M$_{\odot}$, $D_{L}$ is the luminosity 
distance, in this case to M 82  (used in our model computations)
in unit of Mpc, $S_{CO}$ is the $^{12}$CO(1-0) line flux in
units of Jy km s$^{-1}$, and $X$ value is in units of 
10$^{19}$ cm$^{-2}$ [K km s$^{-1}$]$^{-1}$. 

\begin{figure}
\epsscale{1.0}
\plotone{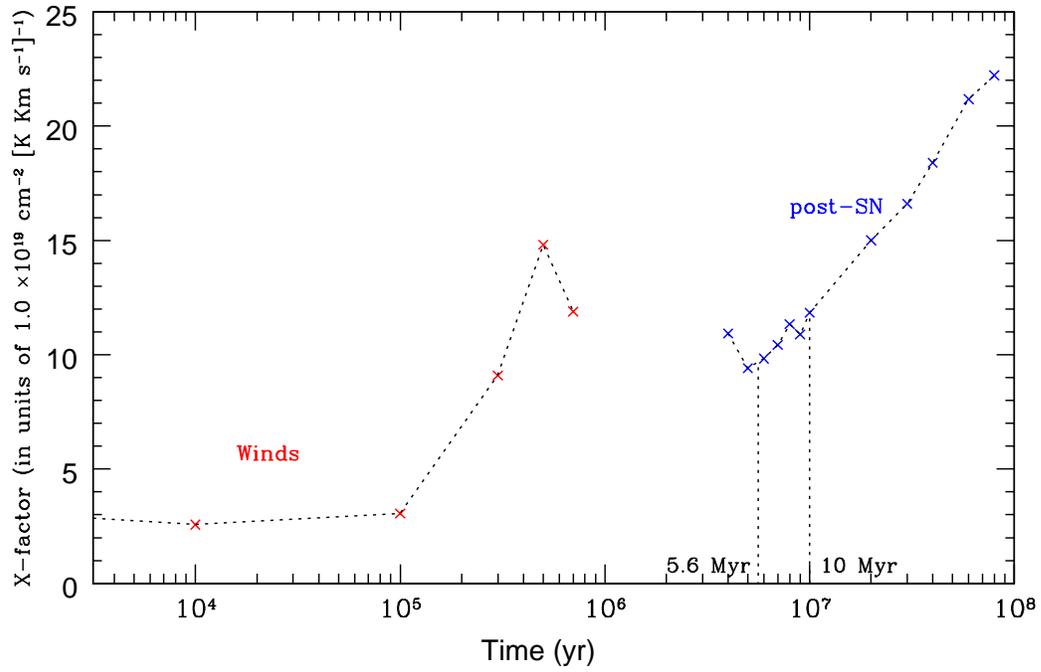}
\caption[Model CO-to-H$_2$ conversion factor $X$ as a function of time]
{Plot of our model CO-to-H$_2$ conversion factor $X$ as a function of time. 
Red cross symbols connect with black dashed curve are for the {\it Winds} 
phase, and blue crosses connected with black dashed curve are for the 
{\it post-SN} phase. The two best fitted ages (5.6 and 10 Myr) are also 
indicated in the plot.
\label{xfact1}}
\end{figure}

In the {\it Winds} phase, the value of $X$ mainly increases with time.
Because the parent GMCs are the dominant sources of $^{12}$CO(1-0) 
line emission during the earlier {\it Winds} phase, and because the 
gas inside the parent clouds is highly excited due to high FUV radiation, 
this results in a progressive decrease in $^{12}$CO(1-0) line 
emission from the GMCs with decreasing GMC mass. On the other hand, 
the compressed dense gas inside the shells is also highly excited,
but the $^{12}$CO(1-0) line emission increases with increasing swept-up 
mass of the shells. Overall the $^{12}$CO(1-0) line emission from the 
shell and GMC ensemble increases with time, and slightly decreases from 
0.5 to 0.7 Myr, while the system H$_2$ mass is fixed at 
1.9 $\times$ 10$^{7}$ M$_{\odot}$. In the {\it post-SN} phase, 
the $X$-factor mainly increases with time, because the $^{12}$CO gas 
is highly excited at early stages of this phase. Although both the 
$^{12}$CO(1-0) line emission and swept-up shell mass $M$(H$_2$) 
increase with time, the increasing rate in $M$(H$_2$) is higher 
than that in the $^{12}$CO(1-0) luminosity between 5 and 80 Myr.
However, the increasing rate in $M$(H$_2$) is lower than that in the 
$^{12}$CO(1-0) luminosity  in a brief interval between 2 and 5 Myr, 
producing a brief decline in the value of $X$ shown in 
Fig.~\ref{xfact1}. The discontinuity between 0.7 and 1 Myr 
corresponds to the transition between {\it Winds} and {\it post-SN} phases. 


The values for the $X$-factor derived from our models for the two best 
fit ages (5.6 and 10 Myr) for the central 1 kpc starburst regions of 
M 82 are 9.5 $\times$ 10$^{19}$ and 1.1 $\times$ 10$^{20}$ cm$^{-2}$ 
[K km s$^{-1}$]$^{-1}$, respectively. These values are comparable
to the empirical values found from the studies of starburst galaxies 
(Weiss et al. 2001, Downes \& Solomon 1998). They also lie between 
those derived for the Galaxy and nearby LIRGs \citep{yao03}. Our model
$X$-factor shows a trend of increasing with time in the {\it post-SN} 
phase, i.e. lower values are associated with more highly excited gas.
This is consistent with the observed results indicated in the above 
references, since starburst galaxies have lower values than those of
the quiescent galaxies like our own Galaxy. 

We also investigated the effect of reducing the upper mass limit on 
the GMC mass spectrum  (i.e. from 10$^7$ M$_{\odot}$ to 
3 $\times$ 10$^6$ M$_{\odot}$ and 10$^6$ M$_{\odot}$) on our model 
$X$-factor as was done for the $^{12}$CO model (Paper II). In Paper II 
we concluded that the model can provide acceptable fits to the data 
only if the dominant initiating starburst clusters are massive, 
at least 5 $\times$ 10$^5$ M$_{\odot}$. Here we find that the 
$X$ value is also very sensitive to the assumed initial upper mass 
limit of the cluster spectrum (and corresponding GMC mass spectrum). 
For the first case with a slightly lower upper limit of GMC mass 
3 $\times$ 10$^6$ M$_{\odot}$ (corresponding to stellar mass of 
7.5 $\times$ 10$^5$ M${\odot}$ with SFE = 25\%; Paper II), we found 
a solution similar to the result presented in Fig.~\ref{xfact1}. The value 
of $X$ is about 15\% and less for the {\it Winds} phase, but it is about
a factor of 1.4 higher at 5.6 Myr and 1.2 higher at 10 Myr for the
{\it post-SN} phase. But for the second case with a ten times or 
more reduction in the upper cutoff of the GMC (and cluster) mass 
spectrum, the value of $X$ starts showing an opposite trend of 
decreasing with time, because in this case the increasing rate 
in $^{12}$CO(1-0) luminosity is faster than the increasing rate 
in $M$(H$_2$) mass swept up by the shells. The value of $X$ is about 
45\% less for the {\it Winds} phase, but it is about a factor of 5 
higher at 5.6 Myr and 2.3 higher at 10 Myr for the {\it post-SN} phase.
Hence, acceptable solutions for modeling the $X$-factor in starburst
galaxies are obtained only when our model system is dominated by 
initiating starburst clusters that are massive, at least 
5 $\times$ 10$^5$ M$_{\odot}$ \citep[e.g.][]{mcc07, kon09}.

The accuracy of the $X$-factor predicted by our model is limited by
several conditions. The CO flux for a given H$_2$ mass depends upon a 
variety of factors which were discussed in Paper II (e.g. the 
dependence on the assumed turbulent velocity, the neglect of 
the CO emission from the ambient ISM). Equally important however is 
that the $X$-factor will depend on the assumed mass 
spectrum for the star clusters (as shown above), and on the assumed 
relation governing the expansion of the bubble driven shells. 
The latter relation governs the amount of H$_2$ mass swept up in a 
given period of time and the hence also the number of star clusters 
required to produce the total observable H$_2$ mass, as discussed also 
in Paper II. 

It is important to understand that a precise value for the $X$-factor 
ultimately relies exclusively on the empirical determinations 
involving careful measurements of CO luminosity and H$_2$ mass. 
What our results do show, however, is that these empirical values 
may be reasonably replicated by a starburst model of the type 
investigated in this study, and that considerable insight regarding 
the causes of the variation from galaxy to galaxy may be obtained 
from the temporal behavior in our model exhibited during the 
expansion of the starburst bubbles/shells.

\section{Conclusions} \label{sum}

As a preliminary approach, we have shown that:

1. By comparing our starburst model with our 
published $^{12}$CO observations of 45 nearby luminous IR galaxies, 
it yields some insight into the relevance of starburst evolution in 
a larger context. Both the model and the data show that the degree of 
CO excitation $r_{31}$ increases with the increasing ratio 
$L_{FIR}$/$M$(H$_2$), and that the frequency distributions of these 
two parameters in both the model prediction and the data are similar. 
This suggests that the observed behavior possibly results from recent 
starbursts in these galaxies observed at different stages of their 
evolution rather than from a wide range of their intrinsic properties 
(e.g. greatly varying degrees of star forming efficiency). This result 
also suggests that the degree of molecular gas excitation is 
a {\it consequence} of star-forming activity, rather than 
a {\it reflection of initial conditions} prior to the starburst, 
as often assumed. The test of the above hypothesis ultimately lies 
in determining the ages of starbursts in many other luminous infrared 
galaxies, most probably by the method of stellar population synthesis. 

2. The CO-to-H$_2$ conversion factor $X$ may be strongly related 
to the starburst phase, because the physical properties of molecular 
gas in an evolving starburst region changes with time as summarized 
before. The model $X$-factor shows a trend of increasing with time in 
the {\it post-SN} phase, i.e. lower values of $X$ are associated 
with more highly excited gas. This is consistent with the observed 
results that starburst galaxies have lower values of $X$-factor 
than those of the non-starburst galaxies. The absolute numerical 
value for $X$-factor derived from our model is sensitive to the 
assumed initial upper mass limit of star clusters spectrum and 
corresponding GMC mass spectrum. In addition, the value will be 
affected to some degree by those factors that affect the age 
prediction described in Paper II.

\acknowledgments

I would like to thank my Ph.D. thesis advisor Ernie Seaquist 
for his guidance and support. I also thank referee for very 
helpful comments. This research was supported by a research grant 
from the Natural Sciences and Engineering Research Council of Canada 
to Ernie Seaquist, and a Reinhardt Graduate Student Travel 
Fellowship from the Department of Astronomy and Astrophysics 
at the University of Toronto.

\end{document}